\documentclass[pra,twocolumn,superscriptaddress,floatfix,amsmath,amsfonts,amssymb,longbibliography]{revtex4-2}%
\usepackage{amsmath,amsfonts,amssymb,color}
\usepackage{amsthm}
\usepackage{leftidx}
\usepackage{graphicx}
\usepackage{subfigure}
\usepackage{xcolor}
\usepackage{dcolumn}
\usepackage{bm}
\usepackage{epstopdf}
\usepackage{epsfig}
\usepackage{physics}
\usepackage{siunitx}
\def\TT{\textcolor{black}}
\def\RB{\textcolor{black}}

\def\GJ{\textcolor{black}}

\begin{document}

	\title{Topological characteristics of gap closing points in nonlinear Weyl semimetals}
	\author{Thomas Tuloup}
	\affiliation{%
		Department of Physics, National University of Singapore, Singapore 117543
	}
	\author{Raditya Weda Bomantara}
	\email{raditya.bomantara@kfupm.edu.sa}
	\affiliation{Centre for Engineered Quantum Systems, School of Physics, University of Sydney, Sydney, New South Wales 2006, Australia}
    \affiliation{Department of Physics, King Fahd University of Petroleum and Minerals, 31261 Dhahran, Saudi Arabia} 
    
	\author{Jiangbin Gong}%
	\email{phygj@nus.edu.sg}
	\affiliation{%
		Department of Physics, National University of Singapore, Singapore 117543
	}
	\date{\today}
	
	%%%%%%%%%%%%%%%%%%%% ABSTRACT %%%%%%%%%%%%%%%%%%%%%%%%
	%\begin{linenumbers}
	
	\vspace{2cm}
	
\begin{abstract}
In this work we explore the effects of nonlinearity on three-dimensional topological phases. Of particular interest are the so-called Weyl semimetals, known for their Weyl nodes, i.e., point-like topological charges which always exist in pairs and demonstrate remarkable robustness against general perturbations. It is found that the presence of onsite nonlinearity causes each of these Weyl nodes to break down into nodal lines and nodal surfaces at two different energies while preserving its topological charge. Depending on the system considered, additional nodal lines may further emerge at high nonlinearity strength. We propose two different ways to probe the observed nodal structures. First, the use of an adiabatic pumping process allows the detection of the nodal lines and surfaces arising from the original Weyl nodes. Second, an Aharonov-Bohm interference experiment is particularly fruitful to capture additional nodal lines that emerge at high nonlinearity.

\end{abstract}

\maketitle
	
\section{Introduction}

Topological phases have been the subject of extensive studies over the past few decades~\cite{Thouless1982,Thouless1983,Kane_and_Mele2005, ExperimentalRealization2007,chen2009experimental,chang2013experimental,Khanikaev2013,Gao2013,Dmitriev2015,lv2015experimental,xu2015discovery,yang2015weyl,Chiu2016,he2017chiral,Volovik2017,Volovik2018,gao2018experimental,imhof2018topolectrical,schindler2018higher,xue2019acoustic,hofmann2020reciprocal}. Among them, Weyl semimetals are of particular interest~\cite{WSMPyrochloreIridates,TopologicalNodalSemimetals,WSMHeterostructure1,TimeReversalInvariantWSM,WSMTransitionMetalMonophosphides,WSMTaAs,WSMTaAs2,EMResponseWSM}. They are three-dimensional (3D) gapless topological phases of matter analogous to graphene, in which electrons behave at low energies as relativistic massless fermions. In a Weyl semimetal, the conduction and valence energy bands touch at a finite number of nodes. These Weyl nodes always come by pair, each carrying opposite chirality and having linear energy dispersion. These Weyl nodes are also robust against general perturbations which only shift their position in quasimomentum space~\cite{Yan2017WSM}, a different behaviour from Dirac cones in graphene where a general perturbation can open a gap. Additionally, Weyl semimetals are an intensely researched topic due to their exotic transport properties~\cite{Wang20173DQuantumHallSemiMetal,Igarashi2017MagnetotransportWSM} such as the chiral anomality effect~\cite{Nielsen1983WeylFermions,Aji2012ABJAnomaly,Son2013ChiralAnomaly,Burkov2014ChiralAnomaly,Gorbar2014ChiralAnomaly,Burkov2015ChiralAnomaly}, a large intrinsic anomalous Hall effect~\cite{Kuroda2017HallEffectWSM}, and perhaps the even more intriguing presence of open Fermi-arc surface states~\cite{WSMPyrochloreIridates,FermiArcSurfaceStates,WSMTaAs2,lv2015experimental,SuYang2015FermiArcs}. These properties rely on the topological nature of the Weyl nodes, which act as monopoles of the Berry curvature. \TT{Weyl semimetals were experimentally realized and the Fermi arcs were observed first in TaAs~\cite{WSMTaAs,WSMTaAs2}, but since then have been realized in many other compounds~\cite{WSMTaP1,WSMTaP2,WSMNbAs}, heterostructures using topological insulator multilayers~\cite{WSMHeterostructure1,WSMHeterostructure2} or even more diverse metamaterials~\cite{Xiao2016ExperimentalWeylMetamaterial,Yang2017ExperimentalWeylPhotonics,Liu2019PhotonicsWSM} and photonic crystals~\cite{Lu2015ExperimentalWeylPhotonics,Noh2017ExperimentalWeylPhotonics}.}

Interaction effect is ubiquitous in nature, and therefore it is of utmost importance to explore its impact and interplay with Weyl points~\cite{InteractingWSMLattice,WSMStrongCoulombLimit,3DTopoPhaseBulkToBoundary,InteractingWeylFermions,EMResponseInteractingWSM,DynamicsElecTransportInteractingWSM}. However, treating interaction effect is challenging due to the exponential increase in the Hilbert space dimension with the system size. Therefore, a mean field approximation is employed to turn a strongly interacting Weyl semimetal into an effective single particle but nonlinear Weyl semimetal\TT{~\cite{Dalfovo1999BEC,Legget2001BEC,Pethick2001BEC,Pitaevskii2003BEC}}. This kind of mean-field approach is especially relevant  to describe the behavior of cold-atom systems such as Bose-Einstein condensates~\cite{Gross1961BEC_GP,Pitaevskii1961BEC_GP,Burger1999DarkSolitonsBEC,Denschlag2000GenerateSolitonsPhaseEngiBEC,Biao2003Review_BEC,Bleu2016ExampleMeanFieldBEC,Watanabe2016ReviewUltracoldAtoms}. Apart from being a mean-field approximation to a strongly interacting system, such nonlinear treatment also naturally appears in photonics setting with the optical Kerr effect, and has received considerable attention~\cite{Lu2013PhotonicCrystalWSM,Lumer2013SelfLocStatesPhotnicTI,Plotnik2013PhotonicGraphene,Morimoto2016TopologicalNatureNLOpticalEffectsSolids,Xin2017KerrEffectNonlinearTopologicalPhotonics,Hadad2018ExampleNonlinearProtectionPossiblePhaseTransition,Smirnova2019NonlinearTopologicalPhotonics}. However, a comprehensive study of nonlinear Weyl semimetals has remained elusive~\cite{WU2011AnomalousMonopoles,NLandTopology}.

In view of the above, we consider here a few variations of a minimal Weyl semimetal lattice model with two Weyl nodes, then investigate the effect of adding an on-site nonlinearity. \TT{Such systems are inherently 3D, and Weyl semimetal phases have been predicted~\cite{Lu2013PhotonicCrystalWSM} and experimentally observed~\cite{Lu2015ExperimentalWeylPhotonics} using 3D photonic crystals based on double-gyroid structures. However, the advent of synthetic dimensions~\cite{Jukic20134DPhotonics,Luo2015Synthetic2DPhotonics,Yuan2016SyntheticDimensions,Ozawa2016SyntheticDimensions,Yuan2018SyntheticDimensions,Lustig2019SyhtheticDimensions,Dutt2020SyntheticDimensions} allows the simulation of some quasimomenta with artificial periodic parameters. As a result, a more convenient experimental setup can in principle be realized in lower dimensional photonic waveguides~\cite{Noh2017ExperimentalWeylPhotonics}, with Kerr nonlinearity arising for sufficiently high optical power.} 

By investigating the systems in momentum space under periodic boundary conditions (PBC), we find that nonlinearity breaks down a Weyl point into nodal lines and nodal surfaces. By evaluating the Chern number of a two-dimensional (2D) surface enclosing these nodal structures, we further confirm that the topological charge of the original Weyl point is preserved. \RB{Interestingly, such a topological charge is now uniformly distributed uniformly throughout the nodal structures rather being concentrated on a point. Moreover, further increase of the nonlinearity strength eventually causes nodal structures originating from two different Weyl points to merge into a variety of exotic shapes that depend on the orientation of the original Weyl points. For some Weyl points' orientation, we further find that} additional nodal lines may emerge at high nonlinear strength. \RB{These nonlinearity-induced nodal lines have zero Chern number and are instead characterized by a quantized Aharonov-Bohm (AB) phase. Such nodal lines can thus be understood as a higher dimensional variation of the nonlinear Dirac cones discovered in Ref.~\cite{NLDC} and will thus be referred to as \emph{nonlinear Dirac lines}}.

To capture the topological properties of the obtained nodal structures, we propose two complementary methods. First, by adapting the theory of adiabatic pumping~\cite{Thouless1982,Thouless1983,Ke2016ThoulessPumping,Lohse2016ThoulessPumping,Nakajima2016ThoulessPumping,Tangpanitanon2016NLTopoPumping,Hayward2018NLPumping,TopoPumpingBlochOscillations} to the nonlinear setting, we show that the pumped charge is a sum of a term proportional to the system's Chern number and another term that depends on the nonlinear strength. In this case, adiabatic pumping can thus be utilized not only to probe the topological charge of a given nodal structure, but also to determine the strength of nonlinearity in the system, which in turn informs the shape of the nodal structure based on our energy band analysis. However, such a method is not suitable to capture the presence of nonlinear Dirac lines, which carry zero Chern number. Motivated by Ref.~\cite{NLDC}, we thus propose to employ an AB interference experiment as an alternative method to probe such nonlinear Dirac lines. \TT{These two methods can respectively be implemented using a 2D array of waveguide and an optical lattice of ultracold atoms.}

This paper is organized as follows. In Sec.~\ref{section:Nonlinear Weyl semimetals}, we introduce our nonlinear Weyl semimetals, and present our results detailing the effect of nonlinearity on the systems' band structure. Major results include the breaking down of Weyl points into nodal lines and nodal surfaces, understanding how their shapes develop as nonlinearity increases, as well as the robustness of these nodal structures against general perturbation. In some of the systems under study, we further reveal the emergence of nonlinear Dirac lines, which are topologically different from the nodal lines arising from the breaking down of Weyl points. In Sec.~\ref{section:Experimental characterization of nodal structures}, we propose two complementary experimental setups to probe the nodal structures in nonlinear Weyl semimetals.
Specifically, by evaluating the particle's displacement in an adiabatic pumping experiment and comparing with the analytical expression derived below, the topological charge and shape of a particular nodal structure that originates from a Weyl point can in principle be obtained. By performing an AB interference experiment, nonlinear Dirac lines can further be probed. In Sec.~\ref{section:Concluding remarks}, we summarize the main findings of this paper and discuss prospects for possible future work.

\section{Nonlinear Weyl semimetals}
\label{section:Nonlinear Weyl semimetals}

\subsection{Nonlinearity as a mean-field approximation to interaction}
\label{section:Nonlinearity as a mean-field approximation to interaction}

We consider a nonlinear version of a two-band Weyl semimetal phase, where two sublattices serve as the pseudospin. It can be described by the stationary Gross-Pitaevskii (GP) equation\TT{~\cite{Gross1961BEC_GP,Pitaevskii1961BEC_GP}}, which writes in momentum space as
\TT{
\begin{equation}
    \label{eqn:GPequation}
     H(\mathbf{k},\ket{\psi(\mathbf{k})}) \ket{\psi(\mathbf{k})} = E(\mathbf{k}) \ket{\psi(\mathbf{k})}
\end{equation}
}
where $\mathbf{k} = (k_x,k_y,k_z)$ is the 3D quasimomentum and
\begin{multline}
    H(\mathbf{k},\ket{\psi(\mathbf{k})}) = h_x(\mathbf{k}) \, \sigma_x + h_y(\mathbf{k}) \, \sigma_y + h_z(\mathbf{k}) \, \sigma_z \\ + g\begin{pmatrix} \left|\psi_{1}(\mathbf{k})\right|^2 & 0 \\ 0 &  \left|\psi_{2}(\mathbf{k})\right|^2 \end{pmatrix}
    \label{eqn:NWSM}
\end{multline}
where $\sigma$'s are the Pauli matrices in the standard representation, \TT{$h_x(\mathbf{k}),h_y(\mathbf{k})$ are the coupling coefficients between the two sublattices, $2 h_z(\mathbf{k})$ is the potential difference between sublattices,} $\ket{\psi(\mathbf{k})} = (\psi_{1}(\mathbf{k}),\psi_{2}(\mathbf{k}))^{T}$ is a Bloch state with two pseudospinor components, and $g$ is the nonlinear strength. \TT{A possible physical realization of this Hamiltonian involves a 2D array of waveguides with the third direction (e.g., the y-axis) being the propagation direction of the light. The propagation direction can be used to represent a quasi-momentum parameter, allowing the realization of a particular 2D snapshot of the 3D Weyl semimetal system at a given parameter (e.g., the quasi-momentum $k_y$). In this case, the nonlinear term in the above Hamilonian represents the Kerr effect naturally arising at high optical powers~\cite{Lumer2013SelfLocStatesPhotnicTI,Plotnik2013PhotonicGraphene,Leykam2016NLKerrEffect}. In an existing experimental study involving waveguide arrays~\cite{Jurgensen2021QuantizedNLThouless}, typical coupling coefficients take values $J \approx 0.06 \, \mathrm{mm}^{-1}$, while the nonlinear strength $g$, assuming nonlinear power normalized to 1 ($|\psi_{1}(\mathbf{k})|^2 + |\psi_{2}(\mathbf{k})|^2 = 1$) take values in the range $0.2J < g < 2.5J$. In this work, all physical variables are assumed to be scaled, and therefore are in dimensionless units.} By defining $\Sigma(\mathbf{k}) = \left|\psi_{2}(\mathbf{k})\right|^2 - \left|\psi_{1}(\mathbf{k})\right|^2$, we can rewrite the nonlinear Hamiltonian in the more convenient form 
\begin{equation}
    H(\mathbf{k},\ket{\psi(\mathbf{k})}) = \frac{g}{2} I_2 + h_x(\mathbf{k}) \, \sigma_x + h_y(\mathbf{k}) \, \sigma_y + (h_z(\mathbf{k}) - \frac{g}{2}\Sigma(\mathbf{k})) \sigma_z. 
    \label{eqn:NWSMSigma}
\end{equation}

In the following, three distinct nonlinear Weyl semimetal systems will be considered. In the first system, which will be referred to as \emph{the perpendicular case}, its Weyl points lie along a line of $h_y=h_z=0$. In the second system, which will be referred to as \emph{the parallel case}, its Weyl points lie along a line of $h_x=h_y=0$. Finally, the third system describes the general case in which the Weyl points lie along a line of $h_x,h_y,h_z\neq 0$. It is worth noting that, in the linear limit, the three systems above are physically equivalent and are unitarily related to one another. Remarkably, in the presence of onsite nonlinearity, the three systems exhibit fundamentally different band structure properties as further elaborated below.

\subsection{Perpendicular case}
\label{section:Perpendicular case}

This part focuses on a nonlinear Weyl semimetal whose Hamiltonian in quasimomentum space is given by Eq.~(\ref{eqn:NWSMSigma}) with $h_x(\mathbf{k}) = (M + \cos{k_x} + \cos{k_y} + \cos{k_z})$, $h_y(\mathbf{k}) = \sin{k_y}$, $h_z(\mathbf{k}) = \sin{k_z}$ and $M = 2$.
The corresponding linear model is a Weyl semimetal phase, exhibiting two Weyl points at $A\equiv (\frac{\pi}{2},\pi,\pi)$ and $B\equiv (-\frac{\pi}{2},\pi,\pi)$. Points A and B respectively have negative and positive chirality, the Chern number around them being -1 and +1.
%\begin{equation}
%    H(\mathbf{k},\ket{\psi(\mathbf{k})}) = h_x(\mathbf{k}) \, \sigma_x + h_y(\mathbf{k}) \, \sigma_y + (h_z(\mathbf{k}) - \frac{g}{2}\Sigma(\mathbf{k})) \sigma_z 
%    \label{eqn:Hperpendicular}
%\end{equation}

Based on the known eigenstate solutions for two-level systems, a stationary state of this Hamiltonian is found to satisfy the self consistency equation (see Appendix~\ref{app:A})
\begin{equation}
    (\frac{g}{2})^2 \Sigma^4 - g h_z \Sigma^3 + [h_x^2 + h_y^2 + h_z^2 - (\frac{g}{2})^2] \Sigma^2 + g h_z \Sigma - h_z^2 = 0,
    \label{eqn:SelfConsistency}
\end{equation}
showing that the nonlinear system can have up to 4 energy bands.

\begin{figure}
  \includegraphics[width=0.9\linewidth]{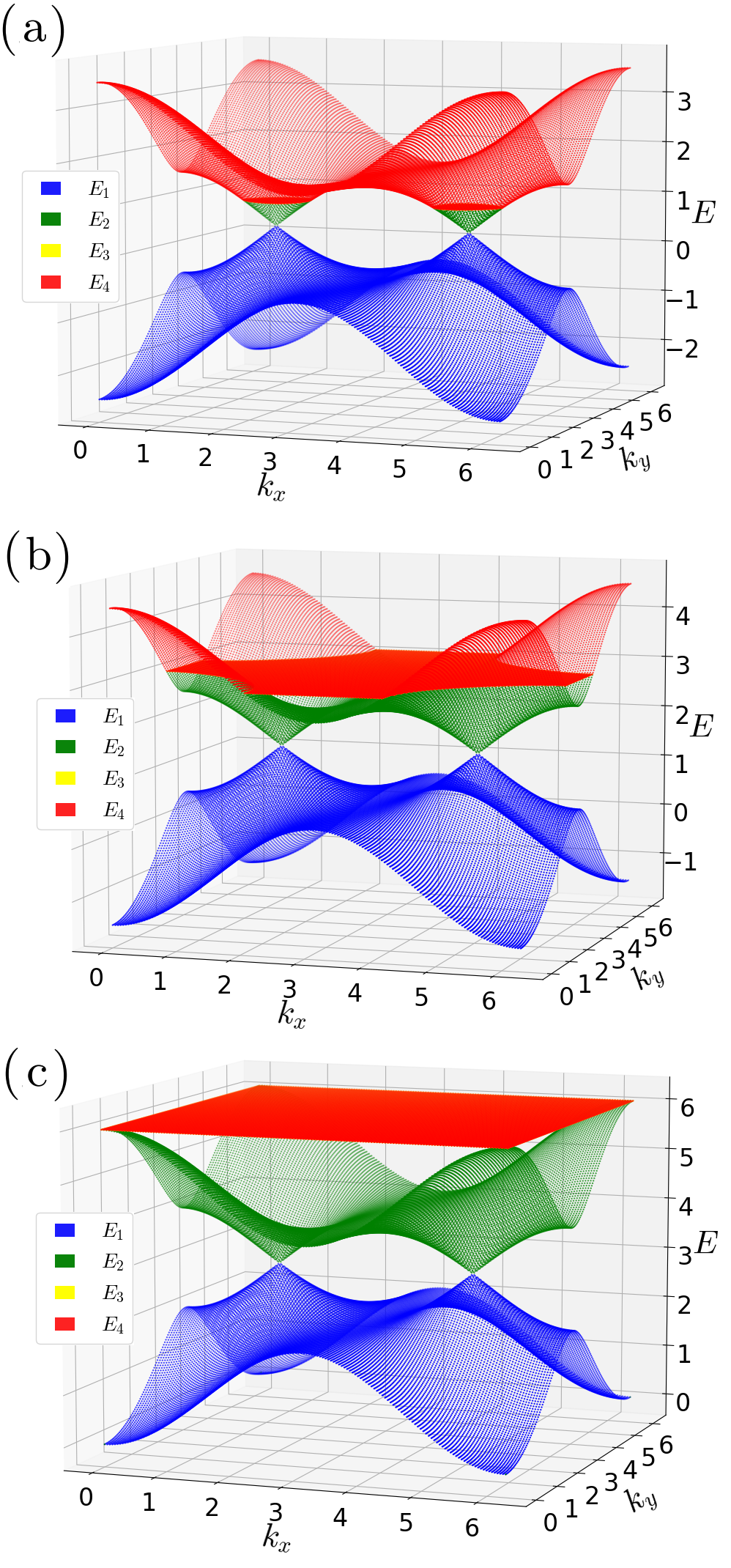}
  \caption{The system's energy bands at different nonlinear strengths \TT{in the perpendicular case}. The third band $E_3$, in yellow, is not visible as it is a flat band, degenerate with $E_4$ at every point in the Brillouin zone where it exists. For all sub-figures we fixed $k_z = \pi$. a) is $g=1$. b) is $g=3$. c) is $g=6$.}
  \label{fig:PerpendicularBandsZ=pi}
\end{figure}

In Fig.~\ref{fig:PerpendicularBandsZ=pi}, we show the system’s band structure at three representative nonlinear strengths. As soon as we add nonlinearity ($g>0$), we notice the following two phenomena. First, while the system can now support up to four energy bands $E_1 \leq E_2 \leq E_3 \leq E_4$, the bands $E_2$ and $E_3$ do not span across the whole the Brillouin zone: they exist only for a small region near the original Weyl points (see green colored bands in Fig.~\ref{fig:PerpendicularBandsZ=pi}-a)). As the nonlinear strength increases, these intermediate bands progressively grow until they merge together as shown in Fig~\ref{fig:PerpendicularBandsZ=pi}-b). Upon further increase in the nonlinear strength, the newly merged band structure grows and eventually spans the entire Brillouin zone as shown in Fig.~\ref{fig:PerpendicularBandsZ=pi} -c).

\begin{figure}
  \includegraphics[width=\linewidth]{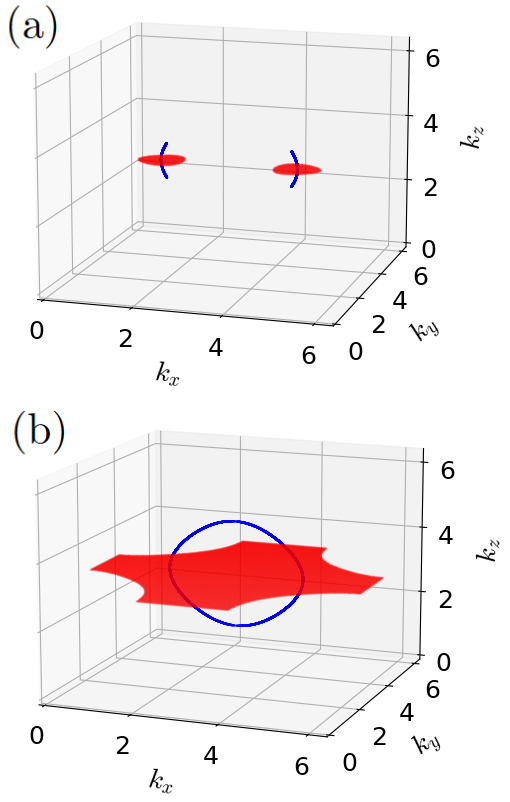}
  \caption{The shape of the nodal lines (blue) and nodal surfaces (red) \TT{of the perpendicular case Hamiltonian} in the 3D Brillouin zone at (a) $g=1$ and (b) $g=3$. The nodal lines (nodal surfaces) correspond to the band touching points at $E_1 = E_2 = \frac{g}{2}$ ($E_3 = E_4 = g$).}
  \label{fig:PerpendicularNodal}
\end{figure}

Second, the proliferation of the number of bands in the system from $2$ to $4$ also yields the splitting of each Weyl point into two different nodal structures separated in energy. The shape of these nodal structures is presented in Fig.~\ref{fig:PerpendicularNodal} for two representative nonlinear strength. Specifically, we find that the lower band touching structure, i.e., $E_1=E_2$, occurs along a 1D line, whereas the upper band touching structure, i.e., $E_3=E_4$, forms a 2D surface. As the nonlinear strength increases, these nodal lines and nodal surfaces eventually merge, as shown in Fig.~\ref{fig:PerpendicularNodal}-b). The analytical derivation of these nodal structures is presented in appendix~\ref{app:A}.

We will now discuss the properties of the above nodal structures in more detail. In a conventional Weyl semimetal such as the corresponding linear Hamiltonian to the one under study, the energy band dispersion around a Weyl point is given by 
\begin{equation}
E_{\pm} = E_0 \pm \sqrt{\kappa_x^2 + \kappa_y^2 + \kappa_z^2}
\end{equation}
where $E_0$ is the energy at the Weyl point and $(\kappa_x,\kappa_y,\kappa_z)$ are small displacements away from the Weyl points along the 3 axis of the 3D Brillouin zone. Accordingly, the dynamics around such a Weyl points are described by the effective Hamiltonian
\begin{equation}
    h_{\text{eff}} = \kappa_x \, \sigma_x + \kappa_y \, \sigma_y + \kappa_z \, \sigma_z,
    \label{Eq:heff}
\end{equation}
which is closely related to Weyl's equation from particle physics, effectively describing a massless fermion, dubbed Weyl fermion. 

\TT{A special property of Weyl points is their robustness against general perturbation, unlike their 2D analogs, the Dirac cones. A general Dirac cone Hamiltonian can be described by the effective Hamiltonian $H_{\textrm{Dirac}} = v_x \, k_x \, \sigma_x + v_z \, k_z \, \, \sigma_z$. where $v,k$ are respectively the group velocities and 2D quasi-momenta. Any perturbation proportional to the third pauli matrix $\sigma_y$ will then open a gap and destroy the Dirac cone. As the effective Hamiltonian presented in Eq.~(\ref{Eq:heff}) contains all three Pauli matrices, a general perturbation will open no gap, and instead will merely shift the position of the Weyl point in the 3D Brillouin zone.}

Using perturbative expansion around the original Weyl points, we are able to study the effect of nonlinearity on the energy bands' dispersion. Let us consider the band touching point $(\frac{\pi}{2},\pi,\pi)$, where we can find degenerate energy solutions $E_1 = E_2 = \frac{g}{2}$ and $E_3=E_4=g$. As detailed in Appendix~\ref{app:B}, at a point $(\frac{\pi}{2}+\kappa_x,\pi+\kappa_y,\pi+\kappa_z)$ near this band touching point, the two lowest and two highest energy bands are respectively modified to become 
\begin{eqnarray} 
\label{eqn:DispersionPerpendicular}
E_{\pm}^{(l)} &=& \frac{g}{2} \pm \sqrt{\kappa_x^2 + \kappa_y^2} \;, \nonumber \\
E_{\pm}^{(h)}&=& g \pm \kappa_z \;
\end{eqnarray}
up to first order in $\kappa_x,\kappa_y$, and $\kappa_z$, where + and - respectively designate the upper energy bands ($E_2$ and $E_4$) and the lowest energy bands ($E_1$ and $E_3$). These energy solutions can be regarded as eigenstates of the effective Hamiltonians
\begin{eqnarray}
\label{eqn:heffPerpendicular}
h_{\text{eff},\pm}^{(l)} &=& \frac{g}{2} I_2 -\kappa_x \, \sigma_x - \kappa_y \, \sigma_y \pm \frac{2 \kappa_z}{g} \sqrt{\kappa_x^2 + \kappa_y^2} \sigma_z\;, \nonumber \\
h_{\text{eff},\pm}^{(h)} &=& \frac{g}{2} I_2 -\kappa_x \, \sigma_x - \kappa_y \, \sigma_y + (-\kappa_z \pm \frac{g}{2}) \sigma_z \;.
\end{eqnarray}
That each energy solution has its own effective Hamiltonian is simply a consequence of the state-dependence of the nonlinear Hamiltonian. While not explicitly shown in this paper, we get similar energy bands dispersion and effective Hamiltonian around the other Weyl point at $(-\frac{\pi}{2},\pi,\pi)$. 

It is particularly interesting to note that the nodal lines only exhibit linear dispersion in the $x$- and $y$-directions, whereas the nodal surfaces only exhibit linear dispersion in the $z$-direction. When viewed together, both nodal lines and nodal surfaces carry the linear dispersion along all three quasimomenta directions as expected from their parent Weyl points. It is also worth noting that while these nodal structures individually only exhibit partial linear dispersion, they inherit the robustness of the original Weyl points \TT{by involving all three Pauli matrices}. Indeed, a perturbation proportional to any of the Pauli matrices will simply displace the locations of the original Weyl points away from $(\pm \frac{\pi}{2},\pi,\pi)$. The same analysis presented in Appendix~\ref{app:B} can then be repeated in the vicinity of the displaced Weyl points to yield exactly the same energy dispersion and effective Hamiltonians as Eqs.~(\ref{eqn:DispersionPerpendicular}) and (\ref{eqn:heffPerpendicular}) respectively. This robustness can also be attributed from the fact that each of these nodal structures carries a topological charge, which can be evaluated through the Chern number of a 2D surface enclosing it.

\subsection{Parallel case}
\label{section:Parallel case}

We now turn to the second system, which is again described by the Hamiltonian of the form Eq.~(\ref{eqn:NWSM}), but with $h_x(\mathbf{k}) = -\sin{k_z}$, $h_y(\mathbf{k}) = \sin{k_y}$, and $h_z(\mathbf{k}) = (M + \cos{k_x} + \cos{k_y} + \cos{k_z})$ and $M=2$. Similarly to the perpendicular case studied in Sec.~\ref{section:Perpendicular case}, at $g=0$, the system exhibits two Weyl points along the $k_x$-axis at $(\frac{\pi}{2},\pi,\pi)$ and $(-\frac{\pi}{2},\pi,\pi)$, which respectively have -1 and +1 chirality.

\begin{figure}
  \includegraphics[width=\linewidth]{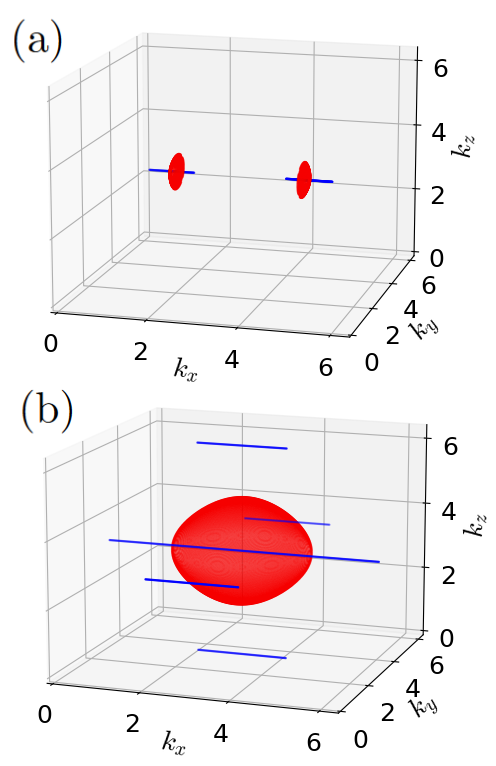}
  \caption{The shape of the nodal lines (blue) and nodal surfaces (red) \TT{of the parallel case Hamiltonian} in the 3D Brillouin zone at (a) $g=1$ and (b) $g=3$. The nodal lines (nodal surfaces) correspond to the band touching points at $E_1 = E_2 = \frac{g}{2}$ ($E_3 = E_4 = g$).}
  \label{fig:ParallelNodal}
\end{figure}

Using the self-consistency equation given in Eq.~(\ref{eqn:SelfConsistency}), we find that as we increase nonlinear strength, the energy bands structure develops in a similar manner as the perpendicular case, where the two full energy bands proliferate into four bands, two of which initially only exist in the vicinity of the Weyl points, splitting the latter into pairs of nodal lines and nodal surfaces (see Fig.~\ref{fig:ParallelNodal}-a)) at $E_1 = E_2 = \frac{g}{2}$ and $E_3 = E_4 = g$ respectively. As the nonlinearity increases, these nodal lines and nodal surfaces grow and eventually merge together as depicted in Fig.~\ref{fig:ParallelNodal}-b).

Despite the aforementioned similarities, two striking differences between the present system and that of Sec.~\ref{section:Perpendicular case} exist. First, the nodal lines are aligned along a direction parallel (perpendicular) to the separation between Weyl points in the present system (the system of Sec.~\ref{section:Perpendicular case}), thus explaining the name ``parallel case" (``perpendicular case") when referring to such a system. Mathematically, taking once again the example of a Weyl point at $(\frac{\pi}{2},\pi,\pi)$, for a point $(\frac{\pi}{2}+\kappa_x,\pi+\kappa_y,\pi+\kappa_z)$ near this original Weyl point, the energies can be pertubatively expanded as
%\begin{equation}
%\label{eqn:DispersionParallel}
%E_{\pm}^{(l)} = \frac{g}{2} \pm \sqrt{\kappa_y^2 + \kappa_z^2},
%\end{equation}

\begin{eqnarray} 
\label{eqn:DispersionParallel}
E_{\pm}^{(l)} &=& \frac{g}{2} \pm \sqrt{\kappa_y^2 + \kappa_z^2} \;, \nonumber \\
E_{\pm}^{(h)}&=& g \pm \kappa_x \;
\end{eqnarray}
up to first order in $\kappa_x,\kappa_y$, and $\kappa_z$. This corresponds to the effective Hamiltonians
\begin{eqnarray}
\label{eqn:heffParallel}
h_{\text{eff},\pm}^{(l)} &=& \frac{g}{2} I_2 + \kappa_z \, \sigma_x - \kappa_y \, \sigma_y \pm \frac{2 \kappa_x}{g} \sqrt{\kappa_y^2 + \kappa_z^2} \sigma_z\;, \nonumber \\
h_{\text{eff},\pm}^{(h)} &=& \frac{g}{2} I_2 + \kappa_z \, \sigma_x - \kappa_y \, \sigma_y + (-\kappa_x \pm \frac{g}{2}) \sigma_z \;.
\end{eqnarray}

\begin{figure}
  \includegraphics[width=\linewidth]{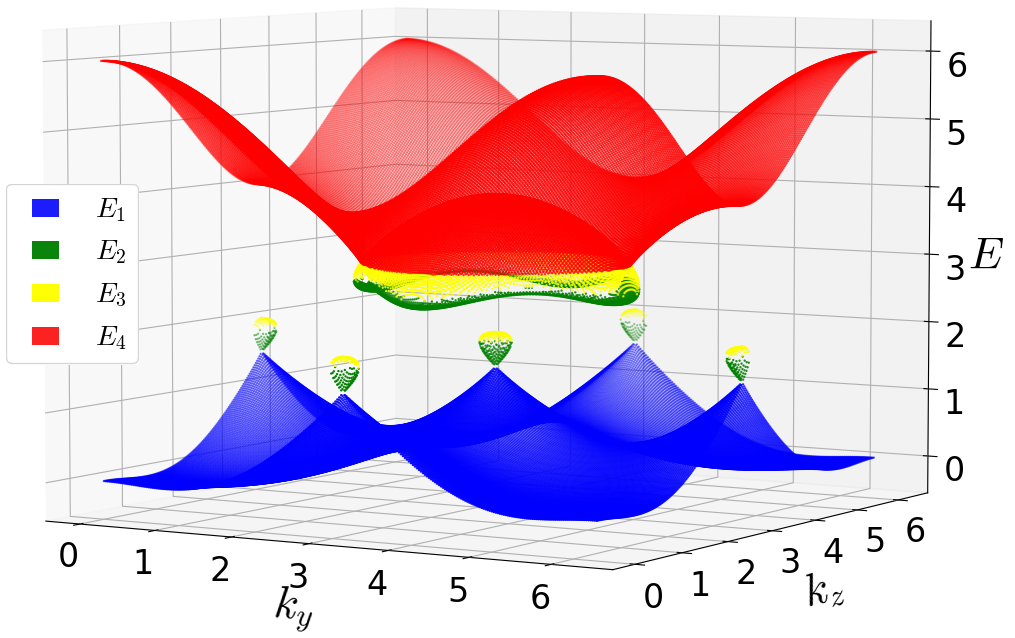}
  \caption{Multiple loop energy bands due to the apparition of nonlinear Dirac lines at high nonlinear strength \TT{in the parallel case}. Parameters are $k_x = \pi$ and $g=3$.}
  \label{fig:ParallelBands}
\end{figure}

Second, as shown in Fig.~\ref{fig:ParallelNodal}-b), the parallel case sports additional nodal lines (nonlinear Dirac lines) at large enough nonlinear strength, not connected to one of the original Weyl points. A snapshot of the system's band structure at a fixed $k_x$, as presented in Fig.~\ref{fig:ParallelBands}, highlights the resemblance of such nodal lines with the nonlinear Dirac cones discovered in Ref.~\cite{NLDC}. In Fig.~\ref{fig:ParallelNodal}-b), we show that these nonlinear Dirac lines are located at $(k_y,k_z) = (0,\pi)$ and $(k_y,k_z) = (\pi,0)$. At an even higher nonlinear strength, an additional nonlinear Dirac line emerges at $(k_y,k_z) = (0,0)$. The energy dispersion around these nonlinear Dirac lines is the same as for the two original nodal lines. For example, for a point $(\pi + \kappa_x , \kappa_y , \pi + \kappa_z)$ near one of these nonlinear Dirac lines, the energy dispersion is
\begin{equation}
\label{eqn:EnergyDispersionNDL}
E^{(\text{NDL})}_{\pm} = \frac{g}{2} \pm \sqrt{\kappa_y^2 + \kappa_z^2},
\end{equation}
while the effective Hamiltonian is 
\begin{equation}
\label{eqn:EffectiveHNDL}
h_{\text{eff},\pm}^{(\text{NDL})} = \frac{g}{2} I_2 + \kappa_z \, \sigma_x + \kappa_y \, \sigma_y \mp \frac{2}{g} \sqrt{\kappa_y^2 + \kappa_z^2} \sigma_z.
\end{equation}

While the nonlinear Dirac lines have the same orientation as the nodal lines that originate from Weyl points, they are different in nature. This is confirmed by evaluating their topological charge, via Chern number calculation through an enclosing surface. In the case of the nodal lines corresponding to a linear Weyl point, appearing as soon as $g > 0$ at $(k_y,k_z)=(\pi,\pi)$, the topological charge is the same as the original Weyl point, -1 for the line going through point $(\frac{\pi}{2},\pi,\pi)$  and +1 for the one going through point $(-\frac{\pi}{2},\pi,\pi)$. On the other hand, repeating the same calculation with respect to a nonlinear Dirac line yields a result of 0, showing that it holds no topological charge. \TT{However, as their effective Hamiltonian still involves all Pauli matrices, nonlinear Dirac lines are equally robust against perturbations. Moreover, as these additional nodal lines are the higher-dimensional generalization of the nonlinear Dirac cones in Ref.~\cite{NLDC}, they can be expected to be associated to another topological invariant in the form of the AB phase associated with two adiabatic paths in the reciprocal space enclosing them, as shown later in this work.}

\subsection{General case}
\label{section:General case}

It is possible to extend the two models presented above to a more general one, by using Pauli transformation in the linear model, before applying the onsite nonlinearity. Taking the perpendicular case Hamiltonian as our starting point, and applying a rotation of angle $\theta$ on $\sigma_x$ and $\sigma_z$ to get $\sigma_x' = \cos{\theta} \, \sigma_x + \sin{\theta} \, \sigma_z$ and $\sigma_z' = \cos{\theta} \, \sigma_z - \sin{\theta} \, \sigma_x$. This again yields the Hamiltonian of Eq.~(\ref{eqn:NWSMSigma}), but with 
%\begin{equation}
%    H_{\theta}(\mathbf{k},\ket{\psi(\mathbf{k})}) = h_x(\mathbf{k},\theta) \, \sigma_x + h_y(\mathbf{k}) \, \sigma_y + (h_z(\mathbf{k},\theta) - \frac{g}{2}\Sigma(\mathbf{k})) \sigma_z 
%    \label{eqn:Hgeneral}
%\end{equation}
\begin{equation}
    \begin{aligned}
    h_x(\mathbf{k},\theta) &= (M + \cos{k_x} + \cos{k_y} + \cos{k_z}) \cos{\theta} - \sin{k_z} \sin{\theta} ,\\
    h_y(\mathbf{k}) &= \sin{k_y} ,\\
    h_z(\mathbf{k},\theta) &= \sin{k_z} \cos{\theta} + (M + \cos{k_x} + \cos{k_y} + \cos{k_z}) \sin{\theta},
    \end{aligned}
\end{equation} 
and $M = 2$. It is immediate to verify that for $\theta = 0$ and $\theta = \frac{\pi}{2}$, we respectively recover the perpendicular case and the parallel case presented earlier. 

\begin{figure}
  \includegraphics[width=\linewidth]{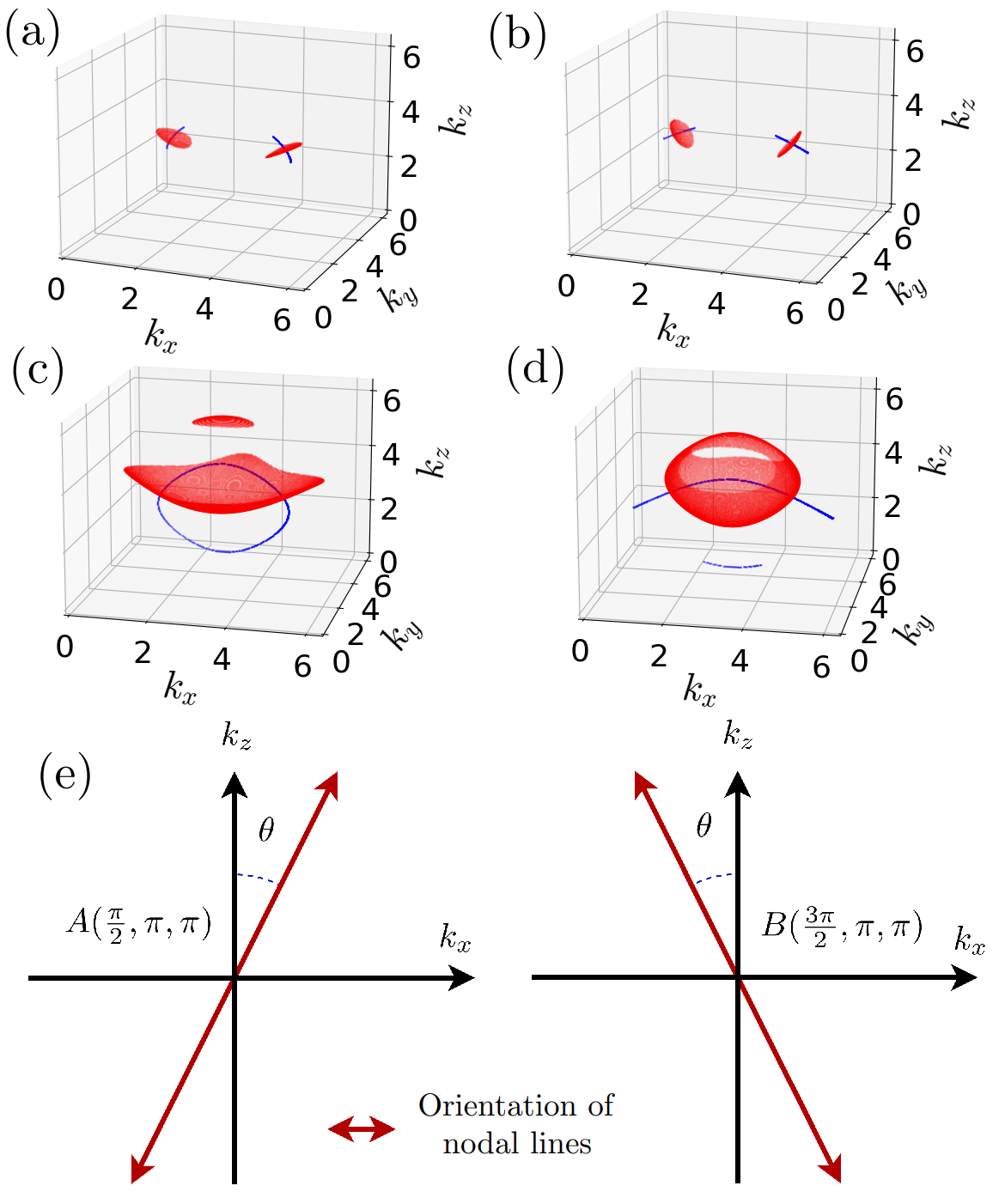}
  \caption{Nodal structures of the general nonlinear Weyl semimetal in the 3D Brillouin zones. In a-d), the 1D nodal curves in blue correspond to the band touching points between the lowest full band and the lower loop band at $E_1 = E_2 = \frac{g}{2}$, and the 2D nodal surfaces, in red, represent the band touching surfaces between the higher loop band and the highest full band at $E_3 = E_4 = g$. In subfigures a) and b), we take $g=1$, and in subfigures c) and d) we take $g=2.5$, a nonlinear strength high enough for the nodal structures to merge. a) and c) are $\theta = \frac{\pi}{6}$. b) and d) are $\theta = \frac{\pi}{3}$. e) is a schematic representation of the nodal lines' direction around both original Weyl points A and B.}
  \label{fig:NodalAngles}
\end{figure}

While such a Pauli rotation does not change the physics of the linear system, it leads to the change in orientation of both the nodal lines and nodal surfaces in the presence of nonlinearity, as shown in Fig.~\ref{fig:NodalAngles}-a) and b). Figure~\ref{fig:NodalAngles}-c) more explicitly highlights the orientation of the nodal lines in the $x-z$ plane in terms of the angle $\theta \notin \{0,\frac{\pi}{2}\}$. Up to first order in $\kappa_x,\kappa_y,$ and $\kappa_z$, the energy dispersions around the two points $A\equiv (\frac{\pi}{2},\pi,\pi)$ and $B\equiv (-\frac{\pi}{2},\pi,\pi)$ are
\begin{equation}
    \begin{aligned}
    E^{(l)}_{A,\pm} &= \frac{g}{2} \pm \sqrt{(-\kappa_x \cos{\theta} + \kappa_z \sin{\theta})^2 + \kappa_y^2} , \\
    E^{(l)}_{B,\pm} &= \frac{g}{2} \pm \sqrt{(\kappa_x \cos{\theta} + \kappa_z \sin{\theta})^2 + \kappa_y^2} ,
    \end{aligned}
\end{equation}
which means that the nodal lines are aligned along:
\begin{equation}
    \text{For A: } \mathbf{k_A} = \begin{pmatrix} \sin{\theta} \\ 0 \\ \cos{\theta} \end{pmatrix} \quad
    \text{For B: } \mathbf{k_B} = \begin{pmatrix} -\sin{\theta} \\ 0 \\ \cos{\theta} \end{pmatrix}.
\end{equation}

The corresponding effective Hamiltonians are
\begin{equation}
\label{eqn:heffLowNodal}
\begin{aligned}
&h_{\text{eff},A,\pm}^{(l)} = \frac{g}{2} I_2 + (-\kappa_x \cos{\theta} + \kappa_z \sin{\theta}) \sigma_x  - \kappa_y \, \sigma_y \\ & \pm \frac{2 (\kappa_z \cos{\theta} + \kappa_x \sin{\theta})}{g} \sqrt{(-\kappa_x \cos{\theta} + \kappa_z \sin{\theta})^2 + \kappa_y^2} \, \sigma_z  \;, \\
&h_{\text{eff},B,\pm}^{(l)} = \frac{g}{2} I_2 + (\kappa_x \cos{\theta} + \kappa_z \sin{\theta}) \sigma_x - \kappa_y \, \sigma_y \\ & \pm \frac{2 (\kappa_z \cos{\theta} - \kappa_x \sin{\theta})}{g} \sqrt{(\kappa_x \cos{\theta} + \kappa_z \sin{\theta})^2 + \kappa_y^2} \, \sigma_z  \;.
\end{aligned}
\end{equation}
Nonlinearity thus provides a means to generate exotic nodal lines along any arbitrary direction in the Brillouin zone.

In the vicinity of the original Weyl points, the nodal surfaces also undergo the same rotation, remaining orthogonal to their corresponding nodal line, with the energy dispersions
\begin{equation}
    \begin{aligned}
    E^{(h)}_{A,\pm} &= g \pm (-\kappa_z \cos{\theta} - \kappa_x \sin{\theta}), \\
    E^{(h)}_{B,\pm} &= g \pm (-\kappa_z \cos{\theta} + \kappa_x \sin{\theta}),
    \end{aligned}
\end{equation}
corresponding to the effective Hamiltonians
\begin{equation}
\label{eqn:heffHighNodal}
\begin{aligned}
h_{\text{eff},A,\pm}^{(h)} &= \frac{g}{2} I_2 + (-\kappa_x \cos{\theta} + \kappa_z \sin{\theta}) \sigma_x - \kappa_y \, \sigma_y \\ & + (-\kappa_z \cos{\theta} - \kappa_x \sin{\theta} \pm \frac{g}{2}) \, \sigma_z\;,  \\
h_{\text{eff},B,\pm}^{(h)} &= \frac{g}{2} I_2 + (\kappa_x \cos{\theta} + \kappa_z \sin{\theta}) \sigma_x - \kappa_y \, \sigma_y \\ & + (-\kappa_z \cos{\theta} + \kappa_x \sin{\theta} \pm \frac{g}{2}) \, \sigma_z \;.
\end{aligned}
\end{equation}
As shown in Fig.~\ref{fig:NodalAngles}-c) and d), the nodal structure eventually \GJ{ends up} merging as nonlinear strength increases, however, the process can wildly differ depending on the choice of angle $\theta$. Although in some cases like $\theta = \frac{\pi}{3}$ it leads to the apparition of nonlinear Dirac lines just like in the parallel case, as shown in Fig.~\ref{fig:NodalAngles}-d), in others such as $\theta = \frac{\pi}{6}$ we witness the creation of additional nodal surfaces not connected to one of the original Weyl points as shown in Fig.~\ref{fig:NodalAngles}-c). Just like the nonlinear Dirac lines, these additional nodal surfaces do not hold any topological charge.

\subsection{Conservation of Fermi-arcs}

One of the most peculiar properties of Weyl semimetals is the presence of Fermi-arc surface states connecting the Weyl points. It is then natural to investigate \GJ{the profiles of such Fermi-arcs in our nonlinear Weyl semimetal model,especially how they change with nonlinearity strength.} For this purpose we consider the system in the general case, i.e., $\theta\in [0,\pi/2]$, but this time we take an infinite slab, infinite along the $x$ and $y$ directions, and finite along the $z$ direction. Using Bloch theorem with the good quantum numbers $k_x,k_y$, we can then study the model as a 1D lattice in real space, along the $z$ direction, whose equations of motion are given in appendix~\ref{app:D}. In the linear case, regardless of $\theta$, the Fermi-arcs consist of degenerate zero energy states along the $x$-axis connecting the two Weyl points. Using an iterative method also described in appendix~\ref{app:D}, we are able to obtain the energy spectrum of the 1D chain in the nonlinear case. We did so under two types of boundary conditions along the $z$ direction, open boundary condition (OBC) and periodic boundary condition (PBC).

\begin{figure}
  \includegraphics[width=\linewidth]{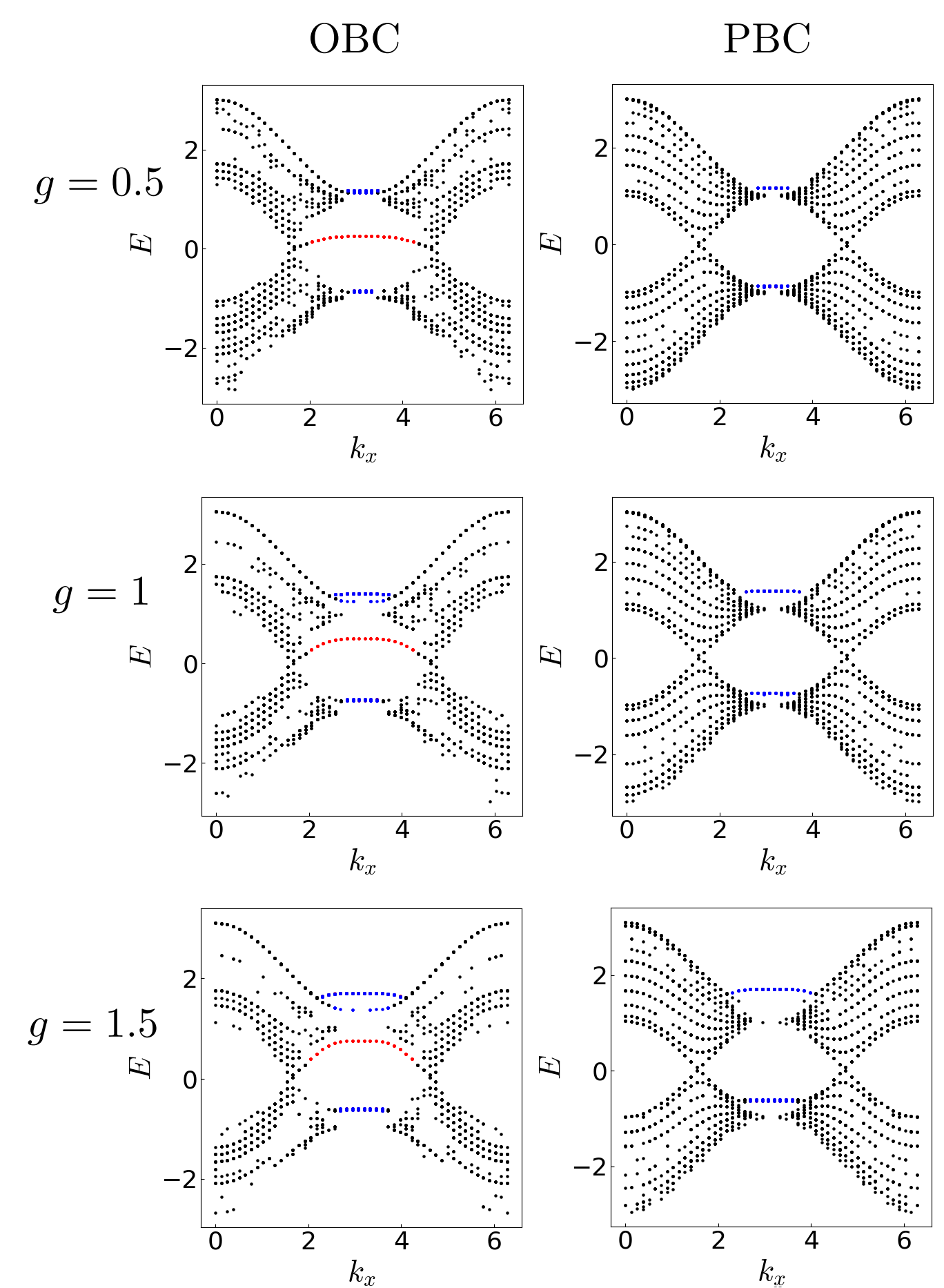}
  \caption{Energy spectrum of the 1D chain along the $z$-axis in real space, assuming PBC and Bloch wave solutions along the $y$ and $z$ axis. Each row corresponds to a different value of the nonlinear strength $g$, while the left and right columns correspond respectively to OBC and PBC along the $x$-axis. \TT{The black dots correspond to delocalized bulk states, while the red dots correspond to strongly localized edge states. The blue dots correspond to strongly localized states located in the bulk, emerging due to the self focusing nonlinearity.} Parameters are $\theta = 0$ \TT{(which corresponds to the perpendicular case)} and $k_y = \pi$, taking $N=20$ unit cells along the $z$-axis. Similar results are observed for other values of $\theta$.}
  \label{fig:FermiArcs}
\end{figure}

The energy spectra for increasing nonlinear strength and both types of boundary conditions are shown in Fig.~\ref{fig:FermiArcs}. The one striking result we observe is the persistence of degenerate edge states, which are however no longer pinned at zero energy, for $\frac{\pi}{2} \leq k_x \leq \frac{3 \pi}{2}$. The fact that these edge states can be observed under OBC but not under PBC shows that they are topological in nature, and is further proof that the Weyl points indeed retain their topological features in the presence of nonlinearity.

\section{Experimental characterization of nodal structures}
\label{section:Experimental characterization of nodal structures}

\subsection{Thouless pumping for detection of nonlinear Weyl points}
\label{section:Thouless pumping for detection of nonlinear Weyl points}

Weyl nodes in Weyl semimetals can already be interpreted as magnetic monopoles for the Berry curvature, and their topological charge can be evaluated by integrating the flux of said Berry curvature through a closed surface enclosing a Weyl node. There is however another way to understand this 2D closed surface of integration; it can be interpreted as describing a band insulator in two dimensions. Indeed, the two parameters needed to describe the enclosing surface can be considered as quasimomenta for the Hamiltonian of a 2D insulator. The corresponding system will be gapped everywhere since the surface is taken to avoid Weyl points. The topology of such 2D insulator can then be characterized by the Chern number, which will be exactly the flux of the Berry curvature through the surface. 

Here, we extend this idea to the 2D surfaces enclosing the nodal structures in Weyl semimetals. For small enough nonlinear stength, it is still possible to define a closed surface around one of the original Weyl points that encloses the full nodal structures, whether they are 1D (nodal lines) or 2D (nodal surfaces).

As an example, let us consider the perpendicular case of Sec.~\ref{section:Perpendicular case}, around point $B(\frac{3 \pi}{2},\pi,\pi)$. As shown on Fig.~\ref{fig:PerpendicularNodal}-a), a straightforward choice for the enclosing surface would be a cylinder centered on point $B\equiv(-\frac{\pi}{2},\pi,\pi)$, with axis of revolution along the $k_z$ direction, and a radius $\rho$ taken large enough to enclose the entire nodal structures. Points on such a cylinder can be described with two parameters $(\phi,z)$ as $k_x = \frac{3 \pi}{2} + \rho \cos{\phi} \, , \, k_y = \pi + \rho \sin{\phi} \, , \, k_z = z$. Due to the periodicity of quasimomenta, this cylinder is actually a torus, and the two parameters $(\phi,z)$ can be immediately taken as the two quasimomenta in a 2D Brillouin zone. This gives us the following Hamiltonian for a 2D Chern insulator
\begin{multline}
    H_{\text{2D}}(k_x,k_y,\ket{\psi(k_x,k_y)}) = \frac{g}{2} I_2 + h_x(k_x,k_y) \, \sigma_x + h_y(k_x) \, \sigma_y \\ + (h_z(k_y) - \frac{g}{2}\Sigma(k_x,k_y)) \sigma_z 
    \label{eqn:H2DChern}
\end{multline}
where
\begin{equation}
    \begin{aligned}
    h_x(k_x,k_y) &= M + \sin{(\rho \cos{k_x})} - \cos{(\rho \sin{k_x})} + \cos{k_y} \\
    h_y(k_x) &= -\sin{(\rho \sin{k_x})} \\
    h_z(k_y) &= \sin{k_y} \\
    \end{aligned}
\end{equation}
and $M=2$.

One convenient way to realize such a system is to actually substitute one of the spatial dimensions for a slow time modulation, realizing a Chern insulator in 1+1 dimensions. \TT{This can be done in a 2D array of waveguides by only exciting a 1D subset of the waveguides, namely a localized Wannier state~\cite{Wannier1397} whose expression in quasi-momentum space characterizes the one spatial dimension of the enclosing surface. The propagation direction, playing the role of time dimension~\cite{Jurgensen2021QuantizedNLThouless}, is then designed so that the excitation we introduced simulates the correct enclosing 2D surface.} In the following, we take $k_y$ as the physical quasimomentum and $k_x = \omega t$ as the time periodic parameter. The resulting system is known as an adiabatic charge pump, and was originally proposed by Thouless~\cite{Thouless1982,Thouless1983}. 
%\begin{multline}
%    H_{\text{Pump}}(k,t,\ket{\psi(k,t}) = h_x(k,t) \, \sigma_x + h_y(t) \, \sigma_y \\ + (h_z(k) - \frac{g}{2}\Sigma(k,t)) \sigma_z 
%    \label{eqn:H1DPump}
%\end{multline}
%where
%\begin{equation}
%    \begin{aligned}
%    h_x(k,t) &= M + \sin{(\rho \cos{\omega t})} - \cos{(\rho \sin{\omega t})} + \cos{k} \\
%    h_y(t) &= -\sin{(\rho \sin{\omega t})} \\
%    h_z(k) &= \sin{k} \\
%    \end{aligned}
%\end{equation} 
%and $M=2$.

In the linear setting, it is known that the average displacement of a particle during one adiabatic cycle in time is equal to the Chern number of the associated 2D Chern insulator\TT{~\cite{Thouless1982,Thouless1983,Simon1983AdiabaticBerryPhase,Niu1984QuantizedTransport,ShortCourseTopologicalInsulators}}. However, nonlinearity is known to nontrivially modify the adiabatic evolution of a system~\cite{NLZP}. In another recent work, we have indeed shown that such modification leads to a nonlinear correction term that generally no longer quantizes the adiabatic pumping result~\cite{NonlinearAdiabaticPumping}. Considering the average displacement of a particle as 
\begin{equation}
    \Delta \left< x \right> = \int \, dt \left< \Bar{v} \right> 
    \label{eqn:Delta}
\end{equation}
with
\begin{equation}
    \label{eqn:vbar}
    \begin{aligned}
        \left< \Bar{v} \right>  &= \frac{1}{2 \pi} \int_{-\pi}^{\pi} \, dk \left< v \right> \\
        &= \frac{1}{2 \pi} \int_{-\pi}^{\pi} \, dk \left< \frac{\partial H}{\partial k} \right>
    \end{aligned}
\end{equation}
where $\left< ... \right>$ represents the average over the state at a given $(k,t)$ \TT{(we set $\hbar=1$ throughout this manuscript)}, we show in Ref.~\cite{NonlinearAdiabaticPumping} that due to nonlinear dynamics, we have
\begin{equation}
    \begin{aligned}
    \Delta \left< x \right> &= \frac{1}{2 \pi} \int  dt \int_{-\pi}^{\pi}  dk  \left[ \mathcal{B}(k,t) + \mathcal{D}(k,t) \right] \;,
    \end{aligned}
    \label{eqn:Displacement}
\end{equation}
where $\mathcal{B}(k,t)$ is the Berry curvature and
\begin{equation}
    \mathcal{D}(k,t) = \frac{g \sin^3{\theta}}{2E - g \sin^2{\theta}} \frac{\partial \phi}{\partial t} \frac{\partial}{\partial k} \left( \frac{\theta}{2} \right) 
    \label{eqn:Drift}
\end{equation}
is a correction to the Berry curvature for the general 2D state $\Psi = (\cos{\frac{\theta(k,t)}{2}},\sin{\frac{\theta(k,t))}{2}} e^{i \phi(k,t)})^{T}$. This correction causes the average displacement to drift away from the integral of the conventional Berry curvature.

\begin{figure}
  \includegraphics[width=\linewidth]{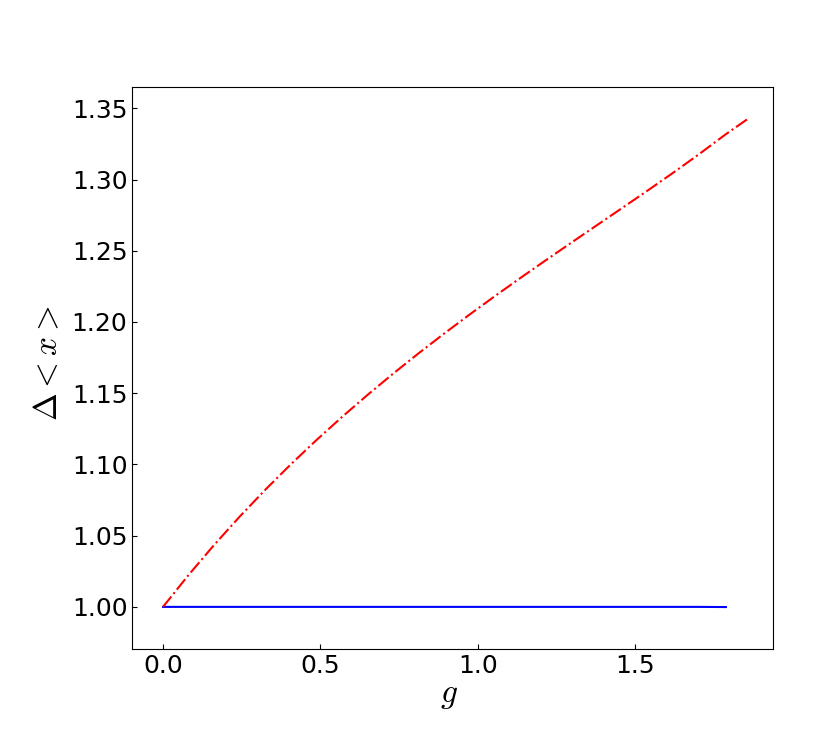}
  \caption{Average displacement of a particle over one adiabatic cycle in the nonlinear Thouless pumping setup. The average displacement without accounting for the additional nonlinear drift is represented by the full blue line, while the red dotted line corresponds to the displacement while accounting for it. Parameters are $\rho = \frac{\pi}{2}$ and $\omega = 10^{-2}$.}
  \label{fig:DisplacementThouless}
\end{figure}

In Fig.~\ref{fig:DisplacementThouless}, we show the average displacement of a particle through numerical simulation of a Thouless pumping in the nonlinear adiabatic pump presented in Eq.~(\ref{eqn:H2DChern}), both with and without taking the additional nonlinear drift into account. When simply integrating the Berry curvature without the nonlinear drift, we expectedly retain a quantized result, equal to the Chern number of the corresponding 2D Chern insulator and the topological charge of the original linear Weyl point. However, this quantization is immediately broken by the addition of the additional drift due to nonlinear dynamics, making it a dead giveaway of the nonlinear nature of the system. \RB{Note also that such a nonlinear drift could serve as a probe to determine $g$ experimentally. To this end, one may carry out the pumping experiment using the specific cylindrical parameterization of Eq.~(\ref{eqn:H2DChern}), then note down the corresponding nonlinear drift, and finally compare it with the numerically obtained value of Fig.~\ref{fig:DisplacementThouless}.} %\TT{Moreover, as the additional nonlinear drift is nearly proportional to $g$, it can be easily used as a probe for the nonlinear strength of the system.}

There are still obstacles to the realization of such a nonlinear Thouless pump, the main one being that the average over quasimomenta $k$ in Eq.~(\ref{eqn:vbar}) is usually obtained in linear systems by taking the initial state to be a Wannier state involving a superposition of Bloch states with all different possible $k$. The superposition principle being lost in the case of nonlinear systems, we are forced to look into other ways to realize this average. One possibility worthy of further investigation is the use of Bloch oscillations to realize this average overall all quasimomenta \cite{TopoPumpingBlochOscillations}. On the other hand, more recent studies highlight that the motion of solitons in a nonlinear Thouless pump is tied to the motion of Wannier states~\cite{Jurgensen2022,mostaan2021quantized}, completely bypassing the need for preparing the latter. However, we warn the reader that the soliton pumping and $k$-averaged pumping generally yield different results. Nonetheless, we show in Ref.~\cite{NonlinearAdiabaticPumping} that both exhibit similar behavior with increase in nonlinearity.

%\comment{(This last sentence seems to be more related on adiabatic pumping result and may therefore be removed. Instead, we can replace it by a follow up sentence to what you previously said in blue, namely, we warn the readers that the soliton pumping and $k$-averaged pumping generally yield different results although they have similar behavior with increase in nonlinearity. We then refer the readers to our other paper for more detail.  )}

\subsection{AB-effect experiment for detection of nonlinear Dirac lines}
\label{section:Experimental proposal}

We showed in the previous section that nonlinear structures connected to original Weyl points of the corresponding linear Weyl semimetal can be detected through an adiabating pumping setup, the displacement being equal to the sum of the topological charge of the original Weyl point and an additional drift giving away the presence of nonlinear effects. However, given that the nonlinear Dirac lines identified in Sec.~\ref{section:Parallel case} do not carry a topological charge, an alternative detection method is deemed necessary. Motivated by the similarity between nonlinear Dirac lines and nonlinear Dirac cones of Ref.~\cite{NLDC}, we propose a means for their detection via an interference setup akin to an AB-effect experiment~\cite{ABInterferometer}. 
\begin{figure}
  \includegraphics[width=\linewidth]{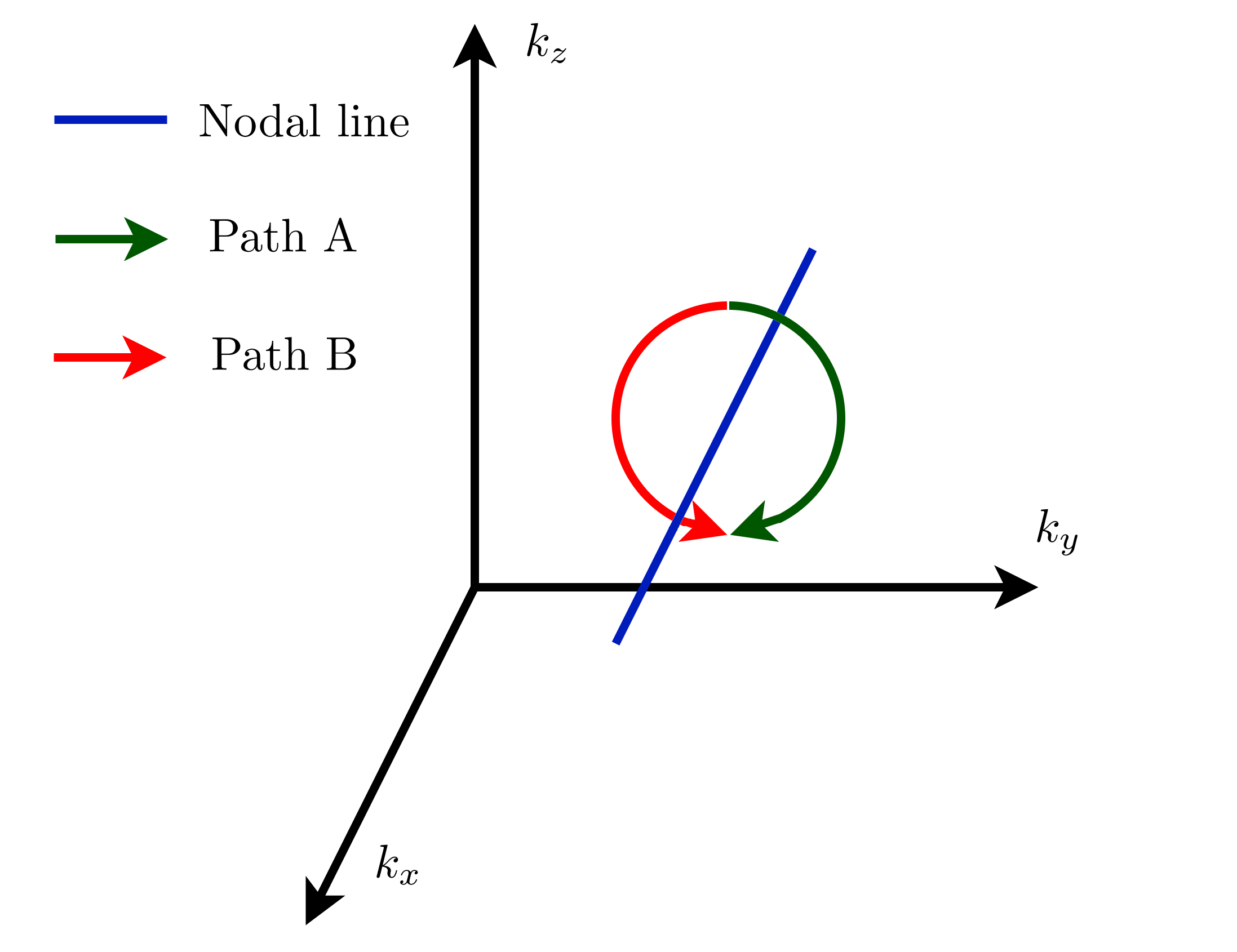}
  \caption{Interfering paths traced adiabatically around a nodal line during the interference experiment. The blue line represents a nodal line, that can be connected to an original Weyl point or not. The green and red arrows respectively represent paths A and B, which describe to semi-circles with opposite senses of rotation, in a plan of fixed $k_x$. The radius of the semi-circles $\rho$ is taken to be very small.}
  \label{fig:ABsetup}
\end{figure}

For this purpose, we study the phase difference between two states adiabatically evolved along two different paths in quasimomentum space. The two interfering paths are shown on Fig.~\ref{fig:ABsetup}, taken to be two semi-circles sharing the same starting point, and going around a nonlinear nodal line in a symmetric manner, with one clockwise and the other one counterclockwise. These paths are designed in the quasimomentum space, \TT{in a plane of fixed $k_x$. During this adiabatic evolution, each state will pick up a geometric phase corresponding to the integral over the path of the Berry connection. The difference of geometric phases amounts to the Berry phase~\cite{BerryPhase1984}, that is alone \emph{non-quantized}. However, the state along each interfering path will also pick up a dynamical phase, constituted of a linear contribution and a nonlinear contribution. Due to the rotational symmetry of the system, the linear contributions remain the same along both paths and cancel out when the phase difference is taken. Meanwhile, the nonlinear contributions to the dynamical phase are of opposite values due to the opposite senses of rotation \RB{and thus lead to non-zero difference. Interestingly, as shown in Ref.~\cite{NLDC}, this non-zero dynamical phase contribution precisely cancels out the nonlinear correction to the Berry phase. As a result, the total phase difference, taking into account both the Berry phase and dynamical phase contributions, yields a quantized value. This result is expected to hold along a contour of constant energy and $m(k_x,k_y)$ for a general nonlinear system, whose effective Hamiltonian in the vicinity of some nodal structures is of the form $H_{\rm eff}= c k_x \sigma_x + c k_y \sigma_y + m(k_x,k_y) c^2 \sigma_z $.}} 
\begin{figure}
  \includegraphics[width=\linewidth]{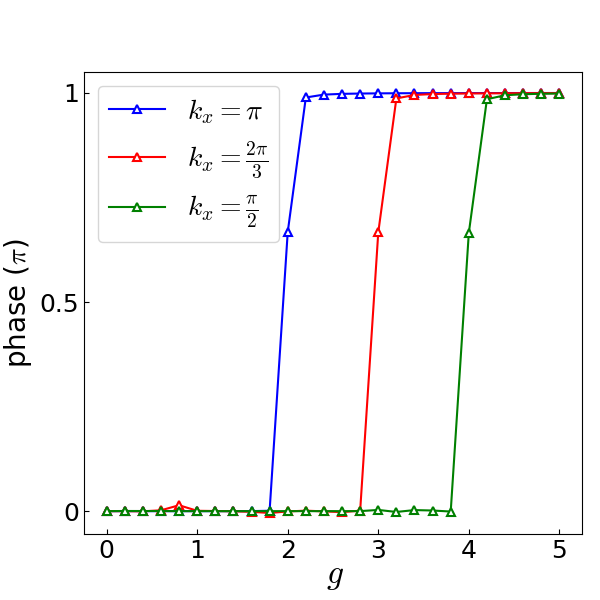}
  \caption{Adiabatic AB phases associated with two interfering paths going around a nonlinear Dirac line with zero topological charge, located at different planes of fixed $k_x$. The system is made to move adiabatically along each path at a frequency $\omega = 10^{-5}$, and each interfering path is a semi-circle of radius $\rho = 10^{-3}$. \TT{The system is taken here in the parallel case, the nodal line considered here being the one at $(k_y,k_z) = (0,\pi)$, but the nodal line at $(k_y,k_z) = (\pi,0)$ yields the same results.}}
  \label{fig:ABDisconnected}
\end{figure}

The results of such an interference setup around a nodal line with zero topological charge are shown on Fig.~\ref{fig:ABDisconnected}. Due to the mutual cancellation of the additional nonlinear contributions to the geometric phase, the results obtained are quantized to a multiple of $\pi$. By fixing the interfering path at a specific $k_x$ value, the AB phase is zero at sufficiently small nonlinearity due to the absence of nonlinear Dirac lines. As the nonlinear strength increases, the nonlinear Dirac lines eventually appear and grow in size until it first goes through the ring at $k_x=\pi$, then at $k_x = \frac{2\pi}{3}$ and later at $k_x = \frac{\pi}{2}$, thus explaining the jump in the AB phase from $0$ to $\pi$ in Fig.~\ref{fig:ABDisconnected}.
\begin{figure}
  \includegraphics[width=\linewidth]{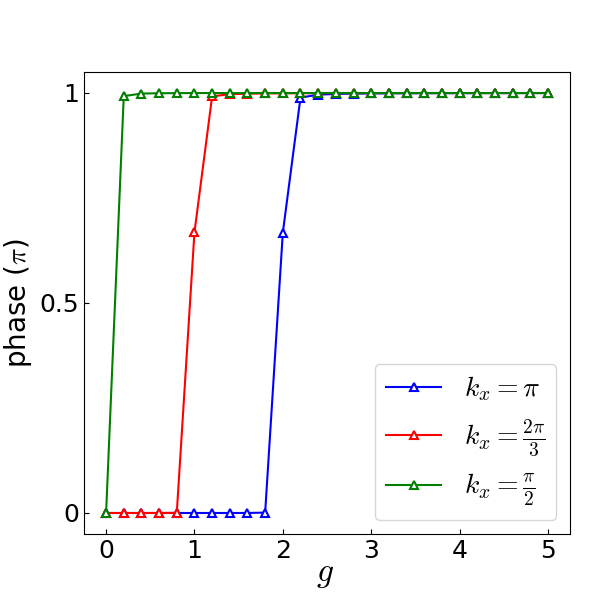}
  \caption{Adiabatic AB phases associated with two interfering paths going around a nonlinear Dirac line with non-zero topological charge, located at different planes of fixed $k_x$. \TT{The system, taken in the parallel case,} is made to move adiabatically along each path at a frequency $\omega = 10^{-5}$, and each interfering path is a semi-circle of radius $\rho = 10^{-3}$.}
  \label{fig:ABConnected}
\end{figure}

One might then wonder what this AB phase would be if the two interfering paths would be instead located around one of the two nodal lines that arise from the breaking down of a linear Weyl point as shown in Fig.~\ref{fig:ParallelNodal}-a), whose topological charge is non-zero. As shown in Fig.~\ref{fig:ABConnected}, a similar conclusion is obtained; it shows zero AB phase when the ring does not go around a nodal line, and an AB phase of $\pi$ as soon as a nodal line goes through it. \TT{Note that the non-quantized data points observed at the jumps between quantized values in Fig.~\ref{fig:ABDisconnected} and Fig.~\ref{fig:ABConnected} are the sign of a topological phase transition, where the AB phase becomes ill-defined at a critical value of nonlinearity strength.} That the AB phase measurement does not make a distinction between nonlinear Dirac lines and nodal lines that originate from a linear Weyl point implies the necessity of using both AB phase measurement and adiabatic pumping as complementary detection schemes.

\TT{The AB phase interferometer experiment can, in principle, be realized in a 2D optical lattice of interacting cold atoms~\cite{ABInterferometer}, realizing a 2D snapshot of the Weyl semimetal at the desired quasi-momentum $k_x$. \RB{Here, the nonlinearity arises from the mean-field description of interaction effect}\TT{~\cite{Dalfovo1999BEC,Legget2001BEC,Pethick2001BEC,Pitaevskii2003BEC}.} Using microwave pulse, a Bose-Einstein condensate is realized in a coherent superposition of spins up and down. Using a spin-dependent force generated by an adiabatically varied magnetic field, it is then possible to trace the two interfering paths in quasi-momentum space. In typical Bose-Einstein condensates, the quotient of the nonlinear interaction term over the typical coupling strength $\frac{g}{J}$~\cite{Dalfovo1999BEC} can go from $\frac{g}{J} \approx 0.5$ with ${}^{7}$Li~\cite{Bradley1997ExperimentalBEC} up to $\frac{g}{J} \approx 10$ with rubidium atoms~\cite{Anderson1995ExperimentalBEC}.}

\section{Concluding remarks}
\label{section:Concluding remarks}

We have earlier elucidated how nonlinearity breaks down a Weyl point into a pair of nodal line and nodal surface, both of which carry the topological charge (Chern number) of the original Weyl point. This then raises a natural question regarding the distribution of this topological charge over the nodal structures. Since nodal lines and nodal surfaces typically hold zero Chern number in linear systems, one may na\"{i}vely think that this topological charge is still localized at the point where the original Weyl point is located. However, our careful analysis below shows otherwise

\begin{figure}
  \includegraphics[width=\linewidth]{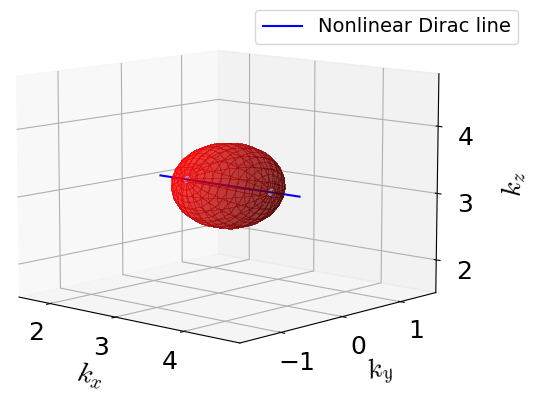}
  \caption{Pierced sphere enclosing a nonlinear Dirac line used for integration of the flux of the Berry curvature. For readability, the holes are larger in the figure than in numerical calculations where their solid angle is $\Omega = \SI{1.57e-4}{\steradian}$.}
  \label{fig:PiercedSphere}
\end{figure}

Specifically, to verify if the Berry curvature was indeed still emitted by a point-like monopole at the original Weyl point's location, we numerically integrated its flux through a ``pierced sphere", centered on the original Weyl point B, and with diminishing radius. This sphere is pierced at two points along the $k_x$ axis, where the nodal lines go through, to avoid the band touching point where the Berry curvature is ill-defined, as shown in Fig.~\ref{fig:PiercedSphere}.

\begin{figure}
  \includegraphics[width=\linewidth]{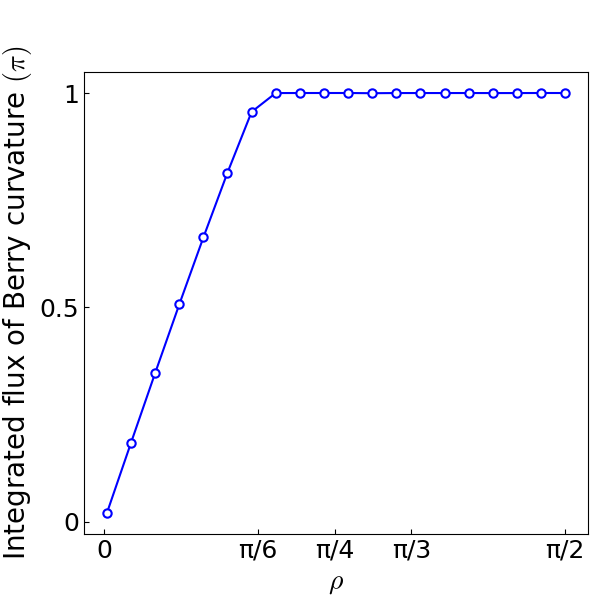}
  \caption{Flux of the Berry curvature through a pierced sphere of radius $\rho$ centered on the original Weyl point $B(\frac{3 \pi}{2},\pi,\pi)$. The sphere is pierced in two points along the line $k_y = k_z = \pi$ to avoid the ill-defined gap closing points. Nonlinear strength is $g=1$, and the solid angle of each hole is $\Omega = \SI{1.57e-4}{\steradian}$.}
  \label{fig:FluxPiercedSphere}
\end{figure}

The numerically integrated flux is shown in Fig.~\ref{fig:FluxPiercedSphere}. As expected, for $\rho \geq \frac{\pi}{6}$, where the entire nodal line is contained in the pierced sphere, the integrated flux is equal to the Chern number with very good precision (up to $10^{-5}$). However, as soon as the pierced sphere does not contain the entire nodal line, i.e. for $\rho < \frac{\pi}{6}$, the integrated flux starts to linearly decrease, all the way to $0$ as the radius goes to $0$. This demonstrates that, contrary to the na\"{i}ve expectation above, the topological charge is in fact uniformly distributed throughout the nodal line. Such a feature is unique to nonlinear systems and has no counterpart in linear nodal-line semimetals.

To conclude, in this work, we have studied in detail the effects of an on-site nonlinearity on the band structure and gap closing points in some Weyl semimetal lattice models exhibiting a single pair of Weyl nodes. One main finding of our work is the breaking down of Weyl nodes into pairs of nodal lines and nodal surfaces, the orientation of which depends on the underlying linear Hamiltonian. At large nonlinearity, additional nodal lines (termed nonlinear Dirac lines) which hold zero topological charge may emerge. We completed this investigation by identifying a proposal for a nonlinear adiabatic pump that allows to access the topological charge of a corresponding nonlinear Weyl point, showing in the process that the additional contribution to the pumping number over an adiabatic cycle due to nonlinear dynamics can be used as an effective probe for the strength of nonlinearity and the shape of the accompanying nodal structure. Finally, we successfully adapted an AB interference setup~\cite{ABInterferometer,NLDC} to detect the presence of nonlinear Dirac lines.

As a possible future direction, the additional two-dimensional nodal surfaces appearing at high nonlinear strength in the general case are yet to be studied and remain elusive, as neither the integration of the flux of the Berry curvature, nor AB interference experiment is able to detect them.
As its interpretation from a mean-field theoric point of view is different from the one-dimensional nodal structures~\cite{WU2011AnomalousMonopoles}, we expect to witness new exciting features. In addition, our observation that nonlinearity deforms the shape of the system's Fermi arcs is also expected to further motivate the study of the interplay between nonlinearity and Fermi arcs in Weyl semimetals. 

Finally, the study of nonlinear effects on other systems is still in its infancy and surely many remains to be discovered. Such systems include higher-order topological systems~\cite{Benalcazar2017,BenalcazarHingeState2017,Langbern2017,LiuHOTP2017,Song2017,Khalaf2018,Schindler2018,Leeli2019HighOrderTopological}, Floquet systems~\cite{Cayssol2013,Gomez2013FloquetBloch,Asboth2014,Grushin2014Floquet,Zhou2014,Bomantra2016Floquet,Linhu2018Floquet}, and non-Hermitian systems~\cite{Longhi2017,Gong2018NonHermitian,MartinezAlvarez2018,Shen2018,Yao2018NonHermitian,Yao2018NonHermitianChern,Ghatak:2019zke,Kawabata2019,Kawabata2019NHSym,Lee2019,Li2019NH,Liu2019NH,Zhou2019NHTopoBands,Ozdemir2019,Okuma2020NonHermitian,Zhao2020NonlinearNonHermitian,Shiqi2021NonlinearNonHermitian}. Even among topological semimetals, many different phases could be investigated, such as nodal line semimetals~\cite{Fang2015NodalLineSemimetals,Fang2015NodalLineSemimetals}, type-II Weyl semimetals~\cite{Soluyanov2015TypeIIWSM,Deng2016TypeIIWSM,Jiang2017TypeIIWSM}, multi-Weyl semimetals~\cite{Fang2012MultiWSM,Umer2021MultiWSM}, and triple-fermion semimetals~\cite{Lv2017TripleFermionSM}.

\begin{acknowledgements}
	{\bf Acknowledgement}: R.W.B is supported by the Australian Research Council Centre of Excellence for Engineered Quantum Systems (EQUS, CE170100009).  J.G. is funded by the Singapore National Research Foundation Grant No. NRF-NRFI2017-04 (WBS No.
R-144-000-378- 281).
\end{acknowledgements}

\appendix

\section{Analytical calculation of band touching structures in nonlinear Weyl semimetal} 
\label{app:A}

We consider a general nonlinear Weyl semimetal whose Hamiltonian in 3D quasimomentum space is given by
\begin{equation}
    H(\mathbf{k},\Sigma) = \frac{g}{2} I_2 + h_x(\mathbf{k}) \, \sigma_x + h_y(\mathbf{k}) \, \sigma_y + (h_z(\mathbf{k}) - \frac{g}{2}\Sigma) \sigma_z 
    \label{eqn:A_Hgeneral}
\end{equation}
where $\Sigma = \abs{\Psi_2}^2 - \abs{\Psi_1}^2$. Based on the known eigenstate solutions for two-level systems, a stationary state $\ket{\Psi} = \begin{pmatrix} \cos{\frac{\theta}{2}} \\ \sin{\frac{\theta}{2}} e^{i \phi} \end{pmatrix}$ of $H(\mathbf{k},\Sigma)$ can be found which satisfies the self consistency equation
\begin{equation}
    \begin{aligned}
    \cos{\theta} &= \frac{h_z - \frac{g}{2}\Sigma}{\sqrt{h_x^2 + h_y^2 + (h_z - \frac{g}{2}\Sigma)^2}} \\
    - \Sigma &= \frac{h_z - \frac{g}{2}\Sigma}{\sqrt{h_x^2 + h_y^2 + (h_z - \frac{g}{2}\Sigma)^2}} \\
    0 &= [h_x^2 + h_y^2 + (h_z - \frac{g}{2}\Sigma)^2]\Sigma^2 - (h_z - \frac{g}{2}\Sigma)^2
    \end{aligned}
    \label{eqn:A_SelfConsistency}
\end{equation}

Hence we get a self-consistency equation for $\Sigma$ in the form of a quartic polynomial, where the self-consistent solutions are the real roots verifying $|\Sigma| \leq 1$.

\begin{equation}
    \label{eqn:A_PolyH1}
    (\frac{g}{2})^2 \Sigma^4 - g h_z \Sigma^3 + [h_x^2 + h_y^2 + h_z^2 - (\frac{g}{2})^2] \Sigma^2 + g h_z \Sigma - h_z^2 = 0.
\end{equation}

This quartic polynomial can be easily factorized in different particular cases. First, if we impose the condition $h_x^2 + h_y^2 = 0$, which in all our models describes a 1D contour, the self-consistency equation can be factorized as
\begin{equation}
    (\Sigma - 1)(\Sigma + 1)\left(\Sigma - \frac{2 h_z}{g}\right)^2 = 0,
\end{equation}
which always has the two simple roots $\Sigma = 1$ and $\Sigma = -1$, with respective energies $E=g-h_z$ and $E=g+h_z$. In addition, if $-\frac{g}{2} \leq h_z \leq \frac{g}{2}$, we also have a double root $\Sigma = \frac{2 h_z}{g}$, with energy $E=\frac{g}{2}$, which corresponds to the 1D nodal lines $E_1 = E_2$ in our models.

Secondly, if we impose the condition $h_z = 0$, which in our models describes a 2D surface, the self-consistency equation can be factorized as
\begin{equation}
     \Sigma^2 \left( \Sigma^2 - \frac{g^2-4(h_x^2 + h_y^2)}{g^2} \right) = 0,
\end{equation}
which always has a double root $\Sigma = 0$ with energies $E = \frac{g}{2} \pm \sqrt{h_x^2 + h_y^2}$. However, if the nonlinearity is strong enough such that $2\sqrt{h_x^2 + h_y^2} \leq \abs{g}$ there are two additional solutions, $\Sigma = \pm \sqrt{\frac{g^2 - 4(h_x^2 + h_y^2)}{g^2}}$, both with energy $E=g$. This corresponds to the 2D nodal surfaces $E_3 = E_4$ in our models.

\section{Energy bands and effective Hamiltonian around original band touching points} 
\label{app:B}

Let us consider a general nonlinear Weyl semimetal described in Eq.~(\ref{eqn:NWSM}). In general, a stationary state of this system verifies
\begin{equation}
    \label{eqn:B_Conditions}
    \begin{aligned}
    \abs{\Psi_1}^2 &= \frac{1}{2} + \frac{h_z}{2(E-g)} \\
    \abs{\Psi_2}^2 &= \frac{1}{2} - \frac{h_z}{2(E-g)} \\
    0 &= E^4 - 3 g E^3 + [\frac{13}{4}g^2 - (h_x^2 + h_y^2 + h_z^2)] E^2 \\ &+ g[-\frac{3g^2}{2} + h_z^2 +4(h_x^2 + h_y^2)] E \\ &+ \frac{g^2(g^2 - h_z^2)}{4} - g^2(h_x^2 + h_y^2).
    \end{aligned}
\end{equation}

For any nonlinear strength, the original Weyl points of our models are always part of the newly formed nodal structures,
%\comment{(Really? I thought they transform into gap closing lines and surfaces?)}
both for the two lowest bands at $E_1 = E_2 = \frac{g}{2}$ and for the two highest bands at $E_3 = E_4 = g$. When considering a small displacement $(\kappa_x,\kappa_y,\kappa_z)$ away from one of these original Weyl points, an energy solution can be obtained perturbatively as 
\begin{equation}
    E = E^{(0)} + E^{(1)} + \cdots
\end{equation}
where $E^{(0)}$ is an energy solution at an original Weyl point, and $E^{(n)}$ is a component of $E$ that is of the order of $\kappa_x^n$, $\kappa_y^n$, or $\kappa_z^n$. By expanding the third line of Eq.~(\ref{eqn:B_Conditions}) up to the first order in $\kappa_x$, $\kappa_y$ and $\kappa_z$, we obtain
\begin{equation}
    \left[4 (E^{(0)})^3 - 9g(E^{(0)})^2 + \frac{13 g^2}{2} E^{(0)} - \frac{3 g^3}{2} \right]  E^{(1)} = 0.
    \label{eqn:B_1storder}
\end{equation}
As we are looking into the energy dispersion near the 1D nodal lines $E_1 = E_2 = \frac{g}{2}$, we take $E^{(0)} = \frac{g}{2}$. We can then verify that in all our models, the term inside the square brackets in Eq.~(\ref{eqn:B_1storder}) is equal to 0, which means that $E^{(1)}$ is not necessarily 0. To find it, we expand the third line of Eq.~(\ref{eqn:B_Conditions}) up to the second order in $\kappa_x$, $\kappa_y$ and $\kappa_z$, which gives
\begin{equation}
    \begin{aligned}
    (E^{(1)})^2& = h_x^2 + h_y^2 \\
    E^{(1)} &= \pm \sqrt{h_x^2 + h_y^2}
    \end{aligned}
\end{equation}
giving the energy dispersions for $E^{(l)}$ in Eq.~(\ref{eqn:DispersionPerpendicular}) and Eq.~(\ref{eqn:DispersionParallel}). Substituting this energy dispersion in the first and second lines of Eq.~(\ref{eqn:B_Conditions}), we get 
\begin{equation}
    \label{eqn:B_Psi1Psi2}
    \begin{aligned}
    \abs{\Psi_1}^2 &\approx \frac{1}{2} - \frac{h_z}{g} \left(1 \pm \frac{2\sqrt{h_x^2 + h_y^2}}{g} \right)  \\
    \abs{\Psi_2}^2 &\approx \frac{1}{2} + \frac{h_z}{g} \left(1 \pm \frac{2\sqrt{h_x^2 + h_y^2}}{g} \right).
    \end{aligned}
\end{equation}
Injecting these expression into Eq.~(\ref{eqn:NWSM}) allows us to rewrite Eq.~(\ref{eqn:GPequation}) as
\begin{equation}
    \begin{pmatrix} \mp \frac{2 h_z \sqrt{h_x^2 + h_y ^2}}{g} & h_x - i h_y \\ h_x + i h_y & \pm \frac{2 h_z \sqrt{h_x^2 + h_y ^2}}{g} \end{pmatrix} \begin{pmatrix} \Psi_1 \\ \Psi_2 \end{pmatrix} = \pm \sqrt{h_x^2 + h_y^2} \begin{pmatrix} \Psi_1 \\ \Psi_2 \end{pmatrix},
\end{equation}
which gives us the first effective hamiltonian $h_{\text{eff},\pm}^{(l)}$ given in Eq.~(\ref{eqn:heffPerpendicular}).

Finally, we investigate the energy dispersion near the 2D nodal line $E_3 = E_4 = g$. Taking $E^{(0)} = g$ and plugging it back in Eq.~(\ref{eqn:B_1storder}) once again yields 0 for the term inside the square brackets, showing that $E^{(1)}$ is not necessarily 0. The same expansion to second order in $\kappa_x$, $\kappa_y$ and $\kappa_z$ gives this time
\begin{equation}
    \begin{aligned}
    (E^{(1)})^2 &= h_z^2 \\
    E^{(1)} &= \pm h_z
    \end{aligned}
\end{equation}
giving the energy dispersion $E^{(h)}$ in Eq.~(\ref{eqn:DispersionPerpendicular}) and Eq.~(\ref{eqn:DispersionParallel}). The substitution in the two first lines of Eq.~(\ref{eqn:B_Conditions}), gives
\begin{equation}
    \label{eqn:B_Psi1Psi2_h}
    \begin{aligned}
    \abs{\Psi_1}^2 &\approx \frac{1}{2} \pm \frac{1}{2}   \\
    \abs{\Psi_2}^2 &\approx \frac{1}{2} \mp \frac{1}{2},
    \end{aligned}
\end{equation}
which once injected in Eq.~(\ref{eqn:NWSM}) allows us to rewrite Eq.~(\ref{eqn:GPequation}) as
\begin{equation}
    \begin{pmatrix} h_z \pm \frac{g}{2} & h_x - i h_y \\ h_x + i h_y & - h_z \mp \frac{g}{2} \end{pmatrix} \begin{pmatrix} \Psi_1 \\ \Psi_2 \end{pmatrix} = \left( \frac{g}{2} \pm h_z \right)  \begin{pmatrix} \Psi_1 \\ \Psi_2 \end{pmatrix},
\end{equation}
giving the effective Hamiltonian $h_{\text{eff},\pm}^{(h)}$ in Eq.~(\ref{eqn:heffPerpendicular}) of the main text.

\section{Real space model and iterative method for nonlinear Fermi-arcs} 
\label{app:D}

In the general case, the equations of motion are given by
\begin{equation}
    \label{eqn:D_EoMBloch}
    \begin{aligned}
    i \frac{d \psi_A}{dt} &= (h_x - i h_y) \psi_B + h_z \psi_A + g |\psi_A|^2 \psi_A \\
    i \frac{d \psi_B}{dt} &= (h_x - i h_y) \psi_A - h_z \psi_B + g |\psi_B|^2 \psi_B
    \end{aligned}
\end{equation}
where
\begin{equation}
    \begin{aligned}
    h_x(\mathbf{k},\theta) &= (M + \cos{k_x} + \cos{k_y} + \cos{k_z}) \cos{\theta} - \sin{k_z} \sin{\theta} ,\\
    h_y(\mathbf{k}) &= \sin{k_y} ,\\
    h_z(\mathbf{k},\theta) &= \sin{k_z} \cos{\theta} + (M + \cos{k_x} + \cos{k_y} + \cos{k_z}) \sin{\theta},
    \end{aligned}
\end{equation} 
and we changed the indices $1,2$ by $A,B$, to better represent the ideas of two sites per unit cell in real space. We consider now a slab, infinite along the $x,y$ directions and finite along the $z$ directions, so that $k_x$ and $k_y$ remain good quantum numbers. By using the Bloch ansatz
\begin{equation}
    \psi_{\alpha} = e^{-i k_z z} \phi_{\alpha,z}
\end{equation}
for $\alpha = A,B$, the equations of motion become
\begin{equation}
    \label{eqn:D_EoMReal}
    \begin{aligned}
    i \frac{d \phi_{A,z}}{dt} &= (M + \cos{k_x} + \cos{k_y})\cos{\theta} \, \phi_{A,z}  \\ &+ [(M + \cos{k_x} + \cos{k_y})\cos{\theta} - i \sin{k_y}] \phi_{B,z}   \\ &- \frac{ie^{i\theta}}{2} \phi_{A,z+1} + \frac{ie^{-i\theta}}{2} \phi_{A,z-1} + \frac{e^{i\theta}}{2} \phi_{B,z+1}  \\ &+ \frac{e^{-i\theta}}{2} \phi_{B,z-1} + g |\phi_{A,z}|^2 \phi_{A,z}, \\
    i \frac{d \psi_{B,z}}{dt} &= -(M + \cos{k_x} + \cos{k_y})\cos{\theta} \, \phi_{B,z} \\ &+ [(M + \cos{k_x} + \cos{k_y})\cos{\theta} + i \sin{k_y}] \phi_{A,z} \\ &+ \frac{ie^{i\theta}}{2} \phi_{B,z+1} - \frac{ie^{-i\theta}}{2} \phi_{B,z-1} + \frac{e^{i\theta}}{2} \phi_{A,z+1}  \\ &+ \frac{e^{-i\theta}}{2} \phi_{A,z-1} + g |\phi_{B,z}|^2 \phi_{B,z} ,
    \end{aligned}
\end{equation}
describing a 1D lattice with two sites per unit cell along the $z$-axis. 
In order to find stationary states of such a nonlinear lattice, we use an iterative method, whose iteration process from a state $\ket{\phi_n}$ to state $\ket{\phi_{n+1}}$ for a given nonlinear, state dependent Hamiltonian $H$ is as follows:
\begin{itemize}
    \item We first compute $H_n = H(\ket{\phi_n})$, the nonlinear state-dependent Hamiltonian evaluated at the state $\ket{\phi_n}$.
    \item We then solve $H_n$ for its eigenstates $\ket{\psi_i}$ with $i = 1,...,2N$.
    \item We finally choose the new state $\ket{\phi_{n+1}}$ as the one eigenstate $\ket{\psi_i}$ closest in distance to the previous $\ket{\phi_n}$. More specifically, we take $\ket{\phi_{n+1}} = \ket{\psi_{i_0}}$ where $\left \| \ket{\phi_n} - \ket{\psi_{i_0}} \right \| \leq \left \| \ket{\phi_n} - \ket{\psi_i} \right \|$ for all $i$. The norm is here defined as $\left\|\ket{\psi}\right\|=\langle \psi | \psi\rangle $.
\end{itemize}
We apply the above-described method until the distance between old and new state is less than an arbitrary $\epsilon$, i.e. $\left\|\ket{\phi_{n}} - \ket{\phi_{n+1}}\right\| < \epsilon$. Throughout this work, we take $\epsilon = 10^{-10}$. 
We used this iterative method to find the energy spectrum of both Hamiltonians $H_{\rm OBC}$ and $H_{\rm PBC}$, corresponding to considering both OBC and PBC in the $z$ direction. For both Hamiltonians, we chose the same trial states as the eigenstates of the linear Hamiltonian ($g=0$) under OBC. This is to make sure that when applying the iterative method for $H_{\rm PBC}$, taking an edge state of the linear model under OBC as a trial state indeed returns a delocalized stationary state, showing that edge states do no exist under PBC even in the nonlinear case.

%\twocolumngrid
\bibliographystyle{apsrev4-2}
%\begin{thebibliography}{99}
\bibliography{Bibliography}
	
\end{document}